\documentclass[12pt]{article}

\evensidemargin
-5mm

\usepackage{epsf}
\usepackage[dvips]{graphicx}

\begin{document}

\begin{center}

{\bfseries SLOPE ANALYSIS FOR ELASTIC PROTON-PROTON AND
PROTON-ANTIPROTON SCATTERING}

\vskip 5mm

V.A. Okorokov$^{\dag}$

\vskip 5mm

{\small {\it Moscow Engineering Physics Institute (State
University), \\ Kashirskoe Shosse 31, Moscow, 115409 Russian
Federation}
\\
$\dag$ {\it E-mail: Okorokov@bnl.gov;~ VAOkorokov@mephi.ru}}
\end{center}

\vskip 5mm

\begin{center}
\begin{minipage}{150mm}
\centerline{\bf Abstract} The diffraction slope parameter is
investigated for elastic proton-proton and proton-antiproton
scattering based on the all available experimental data at
intermediate square of momentum transfer in the main. Energy
dependence of the elastic diffraction slope is approximated by
various analytic functions in a model-independent fashion. The
expanded «standard» logarithmic approximations allow to describe
experimental slopes in all available energy range at qualitative
level reasonably. Various fitting functions differ from each other
both in low energy and very high energy domains. Predictions for
diffraction slope parameter are obtained for elastic proton-proton
scattering at NICA, RHIC and LHC energies, for proton-antiproton
elastic reaction in FAIR energy domain for various approximation
functions at intermediate square of momentum transfer. Difference
of nuclear slopes for proton-antiproton and proton-proton
scattering is investigated in wide momentum transfer range also.
\end{minipage}
\end{center}

\vskip 10mm

\section{Introduction}

In the absence of a pure QCD description of the elastic $pp /
\bar{p}p$ and these large-distance scattering states (soft
diffraction), an empirical analysis based on (almost)
model-independent fits to the physical quantities involved plays
an important role in the extraction of novel information, that can
contribute with the development of useful calculational schemes in
the underlying field theory [1]. Therefore, empirical fits of
energy dependences of global scattering parameters have been used
as a important source of the model-independent information. This
approach for $\sigma_{tot}$ and $\rho$ was recently used in [2].
The third important quantity for nucleon elastic scattering is the
slope parameter. The nuclear slope $B$ for elastic scattering
defined according to the following equation with taking into
account the $t$-dependence:
\begin{equation}
B\left(s,t\right)=\frac{\textstyle
\partial}{\textstyle
\partial t}\left(\ln \frac{\textstyle \partial \sigma\left(s,t\right)}{\textstyle
\partial t}\right), \label{Slope-def}
\end{equation}
is determined experimentally. This quantity is of interest in its
own right, especially for large-distance hadronic physics. On the
other hand the study of $B$ parameter is important, in particular,
for reconstruction procedure of full set of helicity amplitudes
for elastic nucleon scattering [2]. The present status of slope
for elastic $pp$ and $\bar{p}p$ scattering is discussed for
various $|t|$ ranges over the full energy domain.

\section{Slope energy dependence}

We have attempted to describe the energy behaviour of the elastic
nuclear slopes for $pp$ and $\bar{p}p$ reactions. The following
analytic functions are used for fitting of experimental slope
energy dependences here: \vspace*{-0.25cm}
$$
\begin{array}{llr}
\hspace*{3.0cm} B\left(s,t\right)=&
B_{0}\left(t\right)+2a_{1}\left(t\right)\ln\left(s/s_{0}\right),&~~~~~~~~~~~~~~~
(2a)
\label{Fit-1} \\
\hspace*{3.0cm} B\left(s,t\right)=&
B_{0}\left(t\right)+2a_{1}\left(t\right)\ln\left(s/s_{0}\right)+a_{2}\left(t\right)\left
[\ln\left(s/s_{0}\right)\right]^{a_{3}\left(t\right)}, & (2b)
\label{Fit-2}\\
\hspace*{3.0cm} B\left(s,t\right)=&
B_{0}\left(t\right)+2a_{1}\left(t\right)\ln\left(s/s_{0}\right)+a_{2}\left(t\right)
\left(s/s_{0}\right)^{a_{3}\left(t\right)},
& (2c) \label{Fit-3}\\
\hspace*{3.0cm} B\left(s,t\right)=&
B_{0}\left(t\right)+2a_{1}\left(t\right)\ln\left(s/s_{0}\right)+a_{2}\left(t\right)\left
[\ln\left(s/s_{0}\right)\right]^{2}, & (2d) \label{Fit-4}
\end{array}\vspace*{-0.25cm}
$$
where $s_{0}=1$ GeV$^{2}$. These functions were used at study of
slope energy dependence in low $|t|$ domain [3].

Most of experimental investigations as well as theoretical models
are focused on the diffraction region $|t| \simeq 0-0.5$
GeV$^{2}$. The energy dependence of experimental nuclear slope at
low momentum transfer was analyzed in detail recently [3].
Specifically, we have focused on the intermediate $|t|$ domain.
Experimental data are from [4 -- 9]. In the intermediate $|t|$
domain experimental data set is 141 / 85 for $pp / \bar{p}p$
reaction respectively. It seems the mean value of $|t|$ is more
important for separation of experimental results on different
$|t|$ domains than the $|t|$-boundaries of corresponding
measurements. It should be emphasized that the experimental data
for intermediate $|t|$ range were separated on two samples which
corresponded the various parameterization types for differential
cross-section, namely, linear, $\ln\left(d\sigma/dt\right) \propto
\left(-Bt\right)$, and quadratic, $\ln\left(d\sigma/dt\right)
\propto \left(-Bt-Ct^{2}\right)$, function. As known the
measurements of nuclear slope, especially at intermediate $|t|$ do
not form a smooth set in energy, unlike the situation for global
scattering parameters $\rho$ and $\sigma_{tot}$, where there is a
good agreement between various group data [10]. Thus the data
samples for approximations are some smaller because of exception
of points which differ from the other experimental points at close
energies significantly. The maximum fraction of excluded points is
equal 15.3\% at intermediate $|t|$ values.

Figure 1 shows the experimental data and corresponding fits for
slope parameter energy dependence at intermediate $|t|$ for $pp$
and $\bar{p}p$ elastic scattering. The Fig.1a and Fig.1c
correspond the linear approximation of differential cross-section
for for $pp$ and $\bar{p}p$ respectively. Experimental data
obtained at quadratic fit of $d\sigma/dt$ and fitting functions
(2a) -- (2d) are presented at Fig.1b for $pp$ and at Fig.1d for
$\bar{p}p$ collisions. The fitting parameter values are indicated
in Table 1 for various interaction types and for different
$d\sigma/dt$ parameterizations. Usually the fit qualities are
poorer for intermediate $|t|$ values than that for low $|t|$
range. The fitting functions (2a) and (2d) agree with experimental
points qualitatively both for linear (Fig.1a) and quadratic
(Fig.1b) parameterizations of $d\sigma/dt$ for $\sqrt{s} \geq 5$
GeV only. The "expanded" parameterizations (2b), (2c) approximate
experimental data at all energies reasonably. But the (2c)
function shows a very slow growth of slope parameter with energy
increasing at $\sqrt{s} \geq 10^{2}$ GeV (Fig.1a). It should be
stressed that the fitting function (2d) predicts decreasing of the
nuclear slope in high energy domain. Such behavior is opposite the
other fitting function (2a) -- (2c). The $\bar{p}p$ experimental
points from linear parameterization of differential cross-section
are fitted by (2a) at $\sqrt{s} \geq 5$ GeV. The $\bar{p}p$ data
disagreement with Regge-like fitting function very significantly
(Fig.1c). One can see that the experimental data admit the
approximation by (2d) for all energy range but not only for
$\sqrt{s} \geq 5$ GeV. Indeed the fit quality for the first case
much better than for second one. The parameter values are shown in
Table 1 for approximation by (2d) of all available experimental
data. The functions (2c) and (2d) show a very close behaviour at
all energies for $\bar{p}p$ data from linear $d\sigma/dt$
parameterization. These fitting functions have a better fit
quality than (2b). The $\bar{p}p$ data from quadratic
parameterization of $d\sigma/dt$ are fitted by (2a) and (2d)
functions for $\sqrt{s} \geq 5$ GeV only and for all available
energies (Fig.1d). In the last case the fit qualities are much
better and fitting parameters are indicated in the Table 1 for
this energy range namely. As above the functions (2c) and (2d)
show a very close fit quality which is some better than this
parameter for (2b) fitting function. One can see the fit qualities
for (2b) -- (2d) are some better for data from quadratic
parameterization of differential cross-sections than for data from
linear approximation of $d\sigma/dt$. Thus the parameterizations
(2b) -- (2d) agree with data points at qualitative level both for
linear (Fig.1c) and quadratic (Fig.1d) parameterization of
proton-antiproton $d\sigma/dt$ but these fits are still
statistically unacceptable.
\begin{table}
\caption{Fitting parameters for slope energy dependence at
intermediate $|t|$} \label{tab:1} \vspace*{-0.4cm}
\begin{center}
\begin{tabular}{|c|c|c|c|c|c|}
\hline \multicolumn{1}{|l|}{Function} &
\multicolumn{5}{|c|}{Parameter} \\
\cline{2-6} \rule{0pt}{10pt}
 & $B_{0}$, GeV$^{-2}$ & $a_{1}$, GeV$^{-2}$ & $a_{2}$, GeV$^{-2}$ & $a_{3}$ & $\chi^{2}/\mbox{n.d.f.}$ \\
\hline
\multicolumn{6}{|c|}{proton-proton scattering, experimental data for $d\sigma/dt=A\exp\left(-Bt\right)$} \\
\hline
(2a) & $8.15 \pm 0.12$  & $0.169 \pm 0.009$ & --              & --               & $111/29$ \\
(2b) & $10.4 \pm 0.4$   & $0.04 \pm 0.01$   & $-21.5 \pm 1.5$ & $-2.11 \pm 0.12$ & $204/55$ \\
(2c) & $8.8 \pm 0.2$    & $0.13 \pm 0.01$   & $-64 \pm 5$     & $-1.39 \pm 0.06$ & $213/55$ \\
(2d) & $4.06 \pm 0.06$  & $0.9 \pm 0.1$     & $-0.12 \pm 0.02$& --               & $60/28$ \\
\hline
\multicolumn{6}{|c|}{proton-proton scattering, experimental data for $d\sigma/dt=A\exp\left(-Bt-Ct^{2}\right)$} \\
\hline
(2a) & $7.1 \pm 0.2$    & $0.33 \pm 0.02$   & --              & --               & $193/34$ \\
(2b) & $7.9 \pm 0.5$    & $0.26 \pm 0.04$   & $-10 \pm 3$     & $-3.0 \pm 0.8$   & $294/61$ \\
(2c) & $7.5 \pm 0.2$    & $0.29 \pm 0.02$   & $-48 \pm 30$    & $-2.0 \pm 0.4$   & $293/61$ \\
(2d) & $4.0 \pm 0.9$    & $1.0 \pm 0.2$     & $-0.14 \pm 0.04$ & -- & $180/33$ \\
\hline
\multicolumn{6}{|c|}{proton-antiproton scattering, experimental data for $d\sigma/dt=A\exp\left(-Bt\right)$} \\
\hline
(2a) & $11.19 \pm 0.05$ & $0.138 \pm 0.004$ & --              & --               & $1209/27$ \\
(2b) & $\left(-6 \pm 2\right) \cdot 10^{3}$ & $0.46 \pm 0.02$ & $\left(6 \pm 2\right) \cdot 10^{3}$ & $\left(-8 \pm 3\right) \cdot 10^{-4}$ & $950/55$ \\
(2c) & $-1176 \pm 72$   & $6.19 \pm 0.25$   & $1191 \pm 72$   & $\left(-1.12 \pm 0.04\right) \cdot 10^{-2}$ & $719/55$ \\
(2d) & $14.95 \pm 0.14$ & $-0.46 \pm 0.02$  &$0.068 \pm 0.003$& --               & $714/56$ \\
\hline
\multicolumn{6}{|c|}{proton-antiproton scattering, experimental data for $d\sigma/dt=A\exp\left(-Bt-Ct^{2}\right)$} \\
\hline
(2a) & $10.2 \pm 0.2$   & $0.189 \pm 0.011$ & --              & --               & $154/21$ \\
(2b) & $\left(-2\pm 3\right) \cdot 10^{3}$  & $0.46 \pm 0.05$   & $\left(2 \pm 3\right) \cdot 10^{3}$ & $\left(-2 \pm 2\right) \cdot 10^{-3}$ & $121/19$ \\
(2c) & $\left(-7 \pm 2\right) \cdot 10^{2}$ & $4.2 \pm 0.8$     & $\left(7 \pm 2\right) \cdot 10^{2}$ & $\left(-1.4 \pm 0.2\right) \cdot 10^{-2}$   & $108/19$ \\
(2d) & $14.3 \pm 0.6$   & $-0.35 \pm 0.08$  &$0.056 \pm 0.008$& -- & $108/20$ \\
\hline
\end{tabular}
\end{center}
\vspace*{-0.4cm}
\end{table}

One can get a predictions for nuclear slope parameter values for
some facilities based on the results shown above. The $B$ values
at intermediate $|t|$ for different energies of FAIR, NICA, RHIC,
and LHC are shown in the Table 2 based on the fitting parameters
obtained for linear parameterization of $d\sigma/dt$. According to
the fit range function (2a) can predicts the $B$ value for
$\bar{p}p$ scattering in $\sqrt{s} \geq 5$ GeV domain only. As
expected the functions (2c) and (2d) predicted the very close
slope parameter values for FAIR. All fitting functions, especially
(2b) and (2c), predict the close values for nuclear slope in NICA
energy domain. Functions (2a) -- (2c) predict larger values for
$B$ in high-energy $pp$ collisions than (2d) approximation.
Perhaps, the future more precise RHIC results will agree better
with predictions based on experimental data fits under study. The
function (2d) with obtained parameters predicts negative $B$
values at LHC energies.  It should be emphasized that various
phenomenological models predict a very sharp decreasing of nuclear
slope in the range $|t| \sim 0.3 - 0.5$ GeV$^{2}$ at LHC energy
$\sqrt{s}=14$ TeV [11]. Just the negative $B$ value predicted for
LHC at $\sqrt{s}=14$ TeV by (2d) is most close to the some model
expectations [12, 13]. Taking into account recent predictions
based on the fitting functions (2a) -- (2d) for low $|t|$ [3] one
can suggest that the model with hadronic amplitude corresponding
to the exchange of three pomerons [13] describes the nuclear slope
some closer to the experimentally inspired values at LHC energy
both at low and intermediate $|t|$ than other models.

Phenomenological models predicts the zero difference of slopes
$\left(\Delta B\right)$ for proton-antiproton and proton-proton
elastic scattering at asymptotic energies. Here the difference
$\Delta B$ is calculated for each function (2a) -- (2d) with
parameters corresponded $\bar{p}p$ and $pp$ fits: $\Delta
B_{i}\left(s\right)=B^{\bar{p}p}_{i}\left(s\right)-B^{pp}_{i}
\left(s\right),~i=\mbox{2a,...2d}$\footnote{Obviously, one can
suggest various combinations of fitting functions for $\Delta B$
calculations.}. It should be stressed that the equal energy domain
are used in $\bar{p}p$ and $pp$ fits for $\Delta B$ calculations,
i.e. the parameter values obtained by (2d) fitting function of
$\bar{p}p$ data from linear fit of $d\sigma/dt$ for $\sqrt{s} \geq
5$ GeV are used for corresponding $\Delta B$ definition. The
difference $\Delta B_{i}\left(s\right)$ at low $|t|$ values was
calculated based on the recent results from [3]. The energy
dependences of $\Delta B$ are shown at Fig.2a and Fig.2b for low
and intermediate $|t|$ respectively. One can see that the
difference of slopes decreasing with increasing of energy for low
$|t|$ domain (Fig.2a). At present the proton-proton experimental
data at highest available energy 200 GeV don't contradict with
fast (square of logarithm of energy) increasing of slope at high
energies in general case [3]. Such behavior could be agree with
the asymptotic growth of total cross section. But on the other
hand the quadratic function (2d) leads to very significant
difference $\Delta B$ for $\bar{p}p$ and $pp$ scattering in high
energy domain for both low (Fig.2a) and intermediate (Fig.2b)
values of $|t|$. The only Regge-like function (2a) predicts the
decreasing of $\Delta B$ with energy growth at intermediate $|t|$
(Fig.2b). The parameterizations (2b) -- (2d) predict the
decreasing of difference of slopes at low and intermediate
energies and fast increasing of $\Delta B$ at higher energies for
intermediate $|t|$ domain (Fig.2b). As expected the most slow
changing of $\Delta B$ is predicted by Regge-like parametrization
(2a) at asymptotic energies. All fitting functions with
experimentally inspired parameters don't predict the constant zero
values of $\Delta B$ at high energies. But it should be emphasized
that only separate fits were made for experimental data for $pp$
and $\bar{p}p$ elastic reactions here. These results indicate on
the importance of investigations at ultra-high energies both $pp$
and $\bar{p}p$ elastic scattering for many fundamental questions
and predictions connected to the general asymptotic properties of
hadronic physics.
\begin{table}
\caption{Predictions for $B$ based on the functions (2a) -- (2d)
for intermediate $|t|$ domain} \label{tab:2} \vspace*{-0.4cm}
\begin{center}
\begin{tabular}{|c|ccc|cc|cc|ccc|}
\hline \multicolumn{1}{|c|}{Fitting} &
\multicolumn{10}{|c|}{Facility energies, $\sqrt{s}$} \\
\cline{2-11} \rule{0pt}{10pt} function &
\multicolumn{3}{|c|}{FAIR, GeV} & \multicolumn{2}{|c|}{NICA, GeV}
&
\multicolumn{2}{|c|}{RHIC, TeV} & \multicolumn{3}{|c|}{LHC, TeV} \\
\cline{2-11} \rule{0pt}{10pt}
 & 5 & 6.5 & 14.7 & 20 & 25 & 0.2 & 0.5 & 14 & 28 & 42$^*$ \\
\hline (2a) & 12.08 & 12.22 & 12.67 & 10.18 & 10.33 & 11.73 &
12.35 & 14.60 & 15.07 & 15.35 \\
(2b) &  12.46 & 12.23 & 12.03 & 10.39 & 10.49 & 11.10 & 11.29 &
11.88 & 12.00 & 12.07 \\
(2c) & 12.68 & 12.44 & 11.96 & 10.34 & 10.47 & 11.56 & 12.03 &
13.76 & 14.12 & 14.34 \\
(2d) & 12.69 & 12.46 & 11.97 & 10.54 & 10.67 & 9.66 & 7.89 & -5.32 & -9.41 & -12.01 \\
\hline \multicolumn{10}{l}{$^*$\rule{0pt}{11pt}\footnotesize The
ultimate energy upgrade of LHC project [14].}
\end{tabular}
\end{center}
\vspace*{-0.4cm}
\end{table}

\section{Summary}

The main results of this paper are following. The most of all
available experimental data for slope parameter in elastic nucleon
collisions are approximated by different analytic functions. The
situation is more unclear at intermediate $|t|$ values than for
low $|t|$ domain. Only the qualitative agreement is observed
between approximations and experimental points both for $pp$ and
$\bar{p}p$ collisions because of poorer quality of data. But the
suggested "expanded" approximations can be used as a reliable fits
for wide range of momentum transfer at all energies. Predictions
for slope parameter are obtained for elastic proton-proton and
proton-antiproton scattering in energy domains of some facilities.
It seems the phenomenological model with hadronic amplitude
corresponding to the exchange of three pomerons describes the
nuclear slope some closer to the experimental fit inspired values
at LHC energy both at low and intermediate $|t|$ than other
models. The energy dependence of difference of slopes
$\left(\Delta B\right)$ for proton-antiproton and proton-proton
elastic scattering was obtained for fitting functions under study.
The $\Delta B$ parameter shows the opposite behaviours at high
energies for low and intermediate $|t|$ domains (decreasing /
increasing, respectively) for all fitting functions with the
exception of Regge-like one. The last function predicts the slow
decreasing of $\Delta B$ with energy growth. It should be
emphasized that all underlying empirical fitting functions with
experimentally inspired parameter values don't predict the zero
difference of slopes for proton-antiproton and proton-proton
elastic scattering both at low and intermediate $|t|$ for high
energy domain.

\begin{figure}
\begin{center}
\includegraphics[width=14.cm,height=12.5cm]{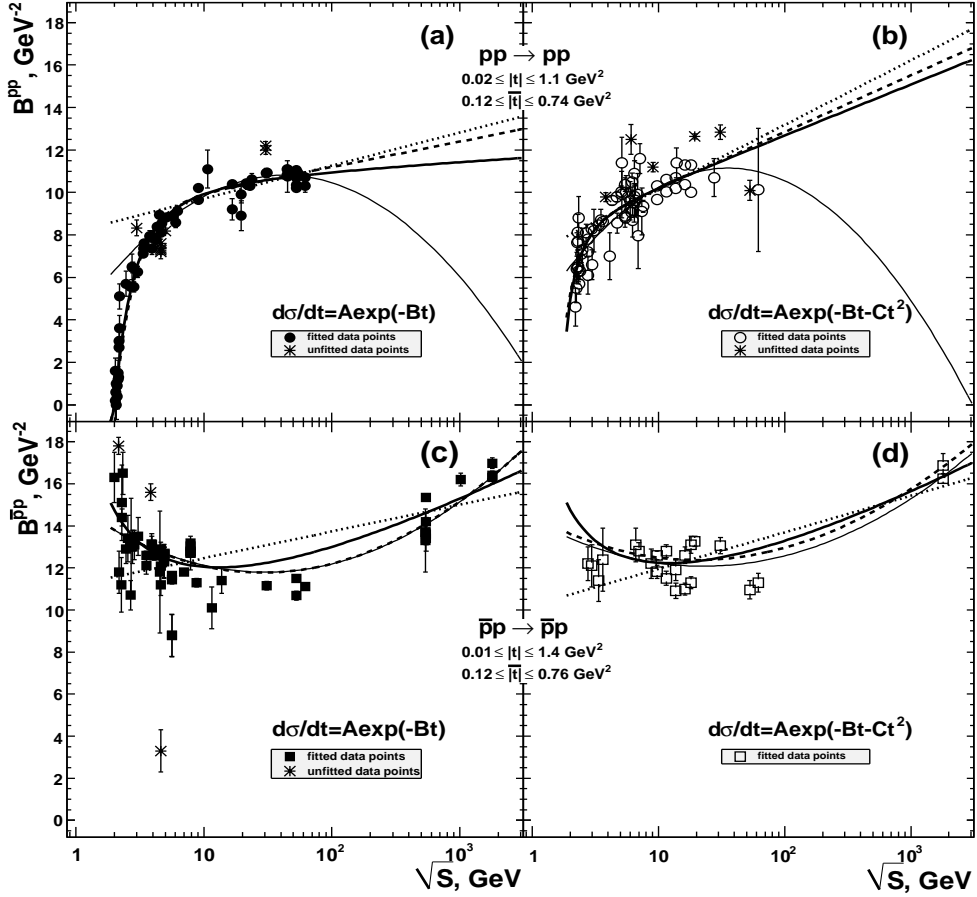}
\end{center}
\vspace*{-0.4cm}
\caption{Energy dependence of the elastic slope
parameters for proton-proton (a,b) and proton-antiproton (c,d)
scattering in intermediate $|t|$ domain for linear (a,c) and
quadratic (b,d) approximation of differential cross-section. The
curves correspond to the fitting functions as following: (2a) --
dot, (2b) -- thick solid, (2c) -- dot-dashed, (2d) -- thin solid.}
\end{figure}
\begin{figure}
\begin{center}
\begin{tabular}{cc}
\mbox{\includegraphics[width=7.5cm,height=7.0cm]{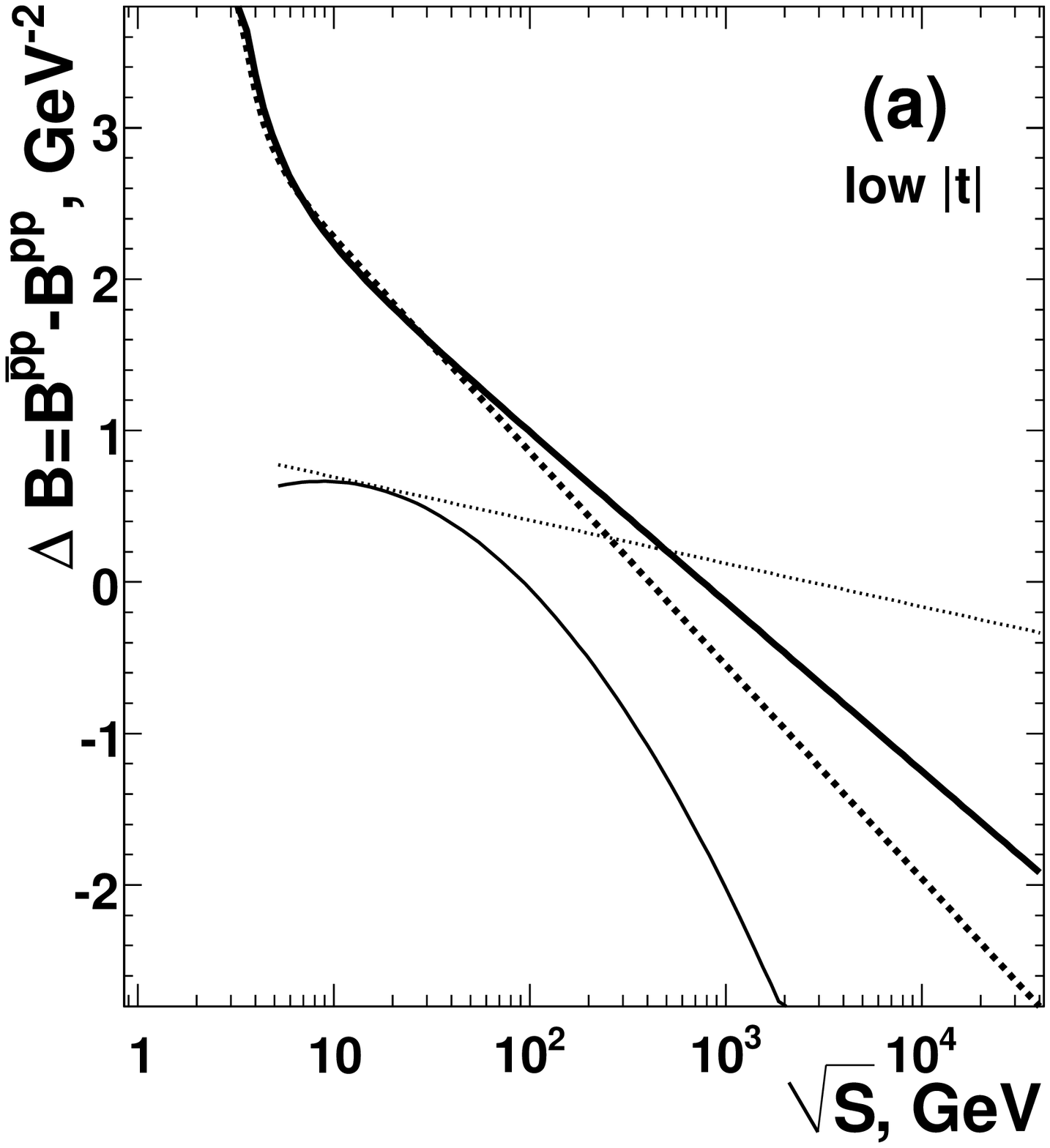}}&
\mbox{\includegraphics[width=7.5cm,height=7.0cm]{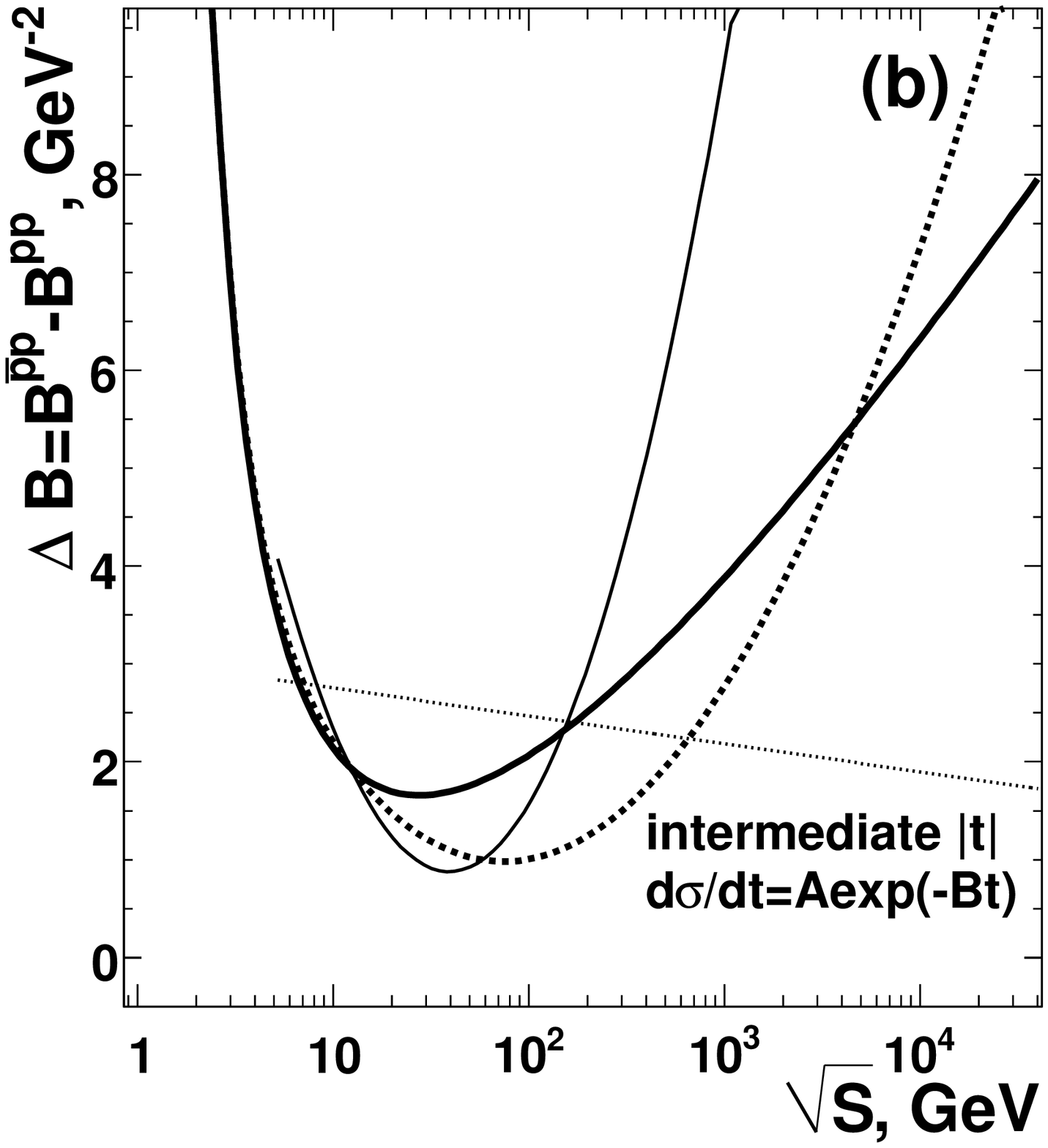}}
\end{tabular}
\end{center}
\vspace*{-0.4cm}
\caption{The energy dependence of the difference
of elastic slopes for proton-antiproton and proton-proton
scattering in low $|t|$ domain (a) and in intermediate $|t|$ range
for linear fit of cross-section (b). The correspondence of curves
to the fit functions is the same as above.}
\end{figure}

\end{document}